\documentclass[cameraready]{Interspeech}
\title{DualTurn: Learning Turn-Taking from Dual-Channel Generative Speech Pretraining}

\author[affiliation={1}]{Shangeth}{Rajaa}

\address{
    $^1$ Anyreach AI
}

\email{shangeth@anyreach.ai}


\begin{document}

\maketitle

\begin{abstract}

Speech-to-speech models handle turn-taking naturally but offer limited support for tool-calling or complex reasoning, while production ASR–LLM–TTS voice pipelines offer these capabilities but rely on silence timeouts, which lead to unnatural turn-taking. We present DualTurn, which narrows this gap through generative pretraining on dual-channel conversational audio. The model generates both speakers' future audio autoregressively, implicitly learning conversational dynamics without any labels, and is then fine-tuned to predict interpretable turn-taking signals that map directly to agent actions. DualTurn monitors both channels continuously, anticipating turn boundaries and producing five agent actions. On standard benchmarks, DualTurn (0.5B) outperforms both VAP on agent action prediction (wF1 0.633 vs.\ 0.389) and a 3.1B audio-text model on word-level turn prediction (AUC 0.930 vs.\ 0.880), while anticipating turn boundaries earlier with fewer interruptions.\end{abstract}

\section{Introduction}

Production voice pipelines built on large language models (LLMs) make use of tool calling and reasoning capabilities but they still depend on silence timeouts for turn-taking, which can cause delayed responses and interruptions. Acoustic features ~\cite{udupa2025nac} and multimodal features~\cite{li2025easyturnintegratingacoustic} also used to address this by recent speech endpointing models, but since they rely on VAD to signal a pause to get triggered, they do not anticipate the speech end, rather are reactive. The generative training of speech-to-speech (S2S) models~\cite{defossez2024moshi, nguyen2023dgslm, roy2026personaplex, wang2025ntppgenerativespeechlanguage} requires predicting the next audio frame given both speaker channels, which allows them to anticipate what either speaker will say next and thus learn turn-taking dynamics implicitly. However, S2S models lack the reasoning and instruction-following capabilities of text LLMs, and their turn-taking abilities cannot be transferred to modular ASR-LLM-TTS pipelines.

Although the turn-taking literature~\cite{sacks1974turntaking, skantze2021turntaking} has advanced considerably, no current model captures the full range of phenomena needed in real-world production pipelines. Text-based models~\cite{ekstedt2020turngpt} predict turn completions from transcripts but ignore the prosodic information that is essential for turn-taking~\cite{gravano2011turntaking, ward2000prosodic}. Audio classifiers~\cite{roddy18_interspeech, raju2023twopass} incorporate prosody but are limited to single-channel binary classification and, without the other speaker's context, can only detect turn end, not the turn action. Backchannels~\cite{ruede2017backchannel}, interruptions, mid-turn pauses, and overlapping speech~\cite{heldner2010pauses} naturally occur in conversational turn-taking. Wang et al.~\cite{wang2024turntaking} use a 3.1B-parameter LLM that leverages both audio and transcript inputs to classify turn shifts, backchannels, and holds, though such models are still too heavy for low-latency deployment. However, just detecting speech end is not enough. The model must also anticipate the speaker's behavior. Voice Activity Projection~(VAP)~\cite{ekstedt2022vap} comes closest, offering a self-supervised, dual-channel model that continuously predicts future voice activity from audio alone. But VAP collapses all phenomena into binary voice-activity probabilities and cannot distinguish backchannels from turn ends; its 5.8M-parameter CPC encoder lacks the capacity for semantic modelling and required additional finetuning to exceed chance on backchannel prediction~\cite{inoue2025backchannel}.

We present DualTurn, a turn-taking component that can be used with ASR-LLM-TTS pipelines using S2S generative pretraining on dual-channel audio. Because DualTurn models both speakers simultaneously, it has access to conversational context that single-channel models miss, such as overlaps, interruptions, and backchannels. It picks up turn-taking and conversational concepts without any labels by learning to predict what either speaker says next. The model encodes the audio with the frozen Mimi neural codec~\cite{defossez2024moshi}, the 0.5B LLM backbone takes in the audio embeddings from the Mimi codec, and predicts six classification signals through MLP heads for each speaker channel. These signals can be combined into agent actions either with a heuristic on the signals or a logistic regression probe. The model can monitor both speakers' channels continuously, running on a single CPU.

DualTurn outperforms VAP across all evaluation setups on Switchboard and otoSpeech. DualTurn outperforms the 3.1B-parameter audio-text fusion model~\cite{wang2024turntaking} on word-level prediction and anticipates turn boundaries before the end of speech with fewer false interruptions. Our main contribution is showing that generative speech pretraining on dual-channel conversational audio, to our knowledge the first use of S2S generative pretraining as a representation-learning stage for explicit turn-taking prediction in modular pipelines, produces representations well-suited to turn-taking signal prediction. From these representations we derive six per-channel signals (turn-ends, mid-turn pauses, speech onsets, backchannels, voice activity, and future activity) that compose into five agent actions. The model runs continuously rather than waiting for VAD to signal a pause.

\section{Method}
\subsection{Architecture}
\begin{figure}[t]

 \centering

 \includegraphics[width=\columnwidth]{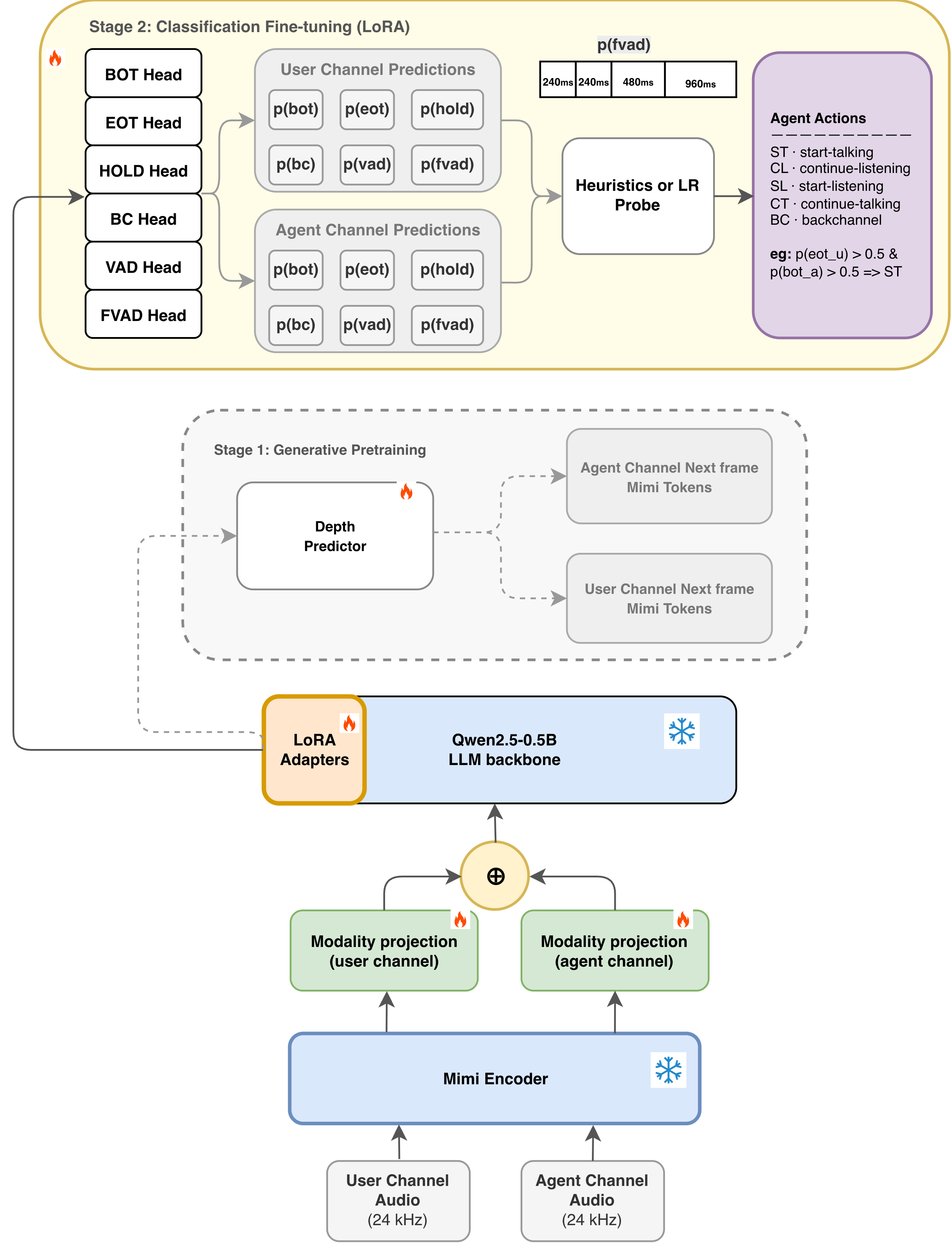}

 \caption{DualTurn architecture. Stage-1 pretrains the backbone and depth predictor (discarded after pretraining) for next speech token prediction; Stage-2 finetunes the twelve classification heads to predict the turn-taking signals.}

 \label{fig:architecture}

\end{figure}

DualTurn uses the Mimi neural codec~\cite{defossez2024moshi} to encode both speakers in a dual-channel audio stream, which converts each channel's 24\,kHz waveforms to 8 RVQ codebook sequences at 12.5\,frames$\cdot$s$^{-1}$. We use the continuous encoder embeddings (512-dim per channel) rather than discrete codebook indices. Each channel's features pass through a channel-specific MLP; the two representations are concatenated and fed to Qwen2.5-0.5B~\cite{qwen2025}, trained in two stages as described below. Twelve lightweight classification heads (six per channel) are attached to the backbone's final hidden state. Sparse signals (EOT, HOLD, BOT, BC) are predicted by two-layer MLP heads with GELU activation and dropout, while the dense signals (VAD, FVAD) require only linear projections. The FVAD heads each output four horizon values, bringing the total to 18 scalar outputs across the 12 heads (the two FVAD heads each produce four horizon values; all others produce one). (Figure~\ref{fig:architecture}). During inference, the model handles streaming audio from both speaker channels and makes predictions with 240\,ms stride (3 audio frames per inference) with KV-caching. The latency of base unquantized model is around 78ms on CPU and 27ms on A100 GPU.

\subsection{Stage-1: Generative Speech Pretraining}

The backbone is pretrained to autoregressively predict both speakers' next audio token simultaneously. The lightweight depth predictor (${\sim}$10.6M parameters) generates the next-frame RVQ codebook tokens per channel from the backbone's output. Once Stage-1 is complete, the depth predictor is discarded and only the representation backbone is retained for Stage-2 training.

The generative training loss forces the model to learn and understand the semantic, prosodic information and interaction patterns, which can then be useful for turn-taking predictions in Stage-2 finetuning.

\subsection{Stage-2: Turn-Taking Signal Prediction}
Six turn-taking signals per channel are predicted from the pretrained backbone in Stage-2. The labels for these signals are derived entirely using self-supervised voice activity alignment without any manual annotation (Table~\ref{tab:signals}).

\begin{table}[t]

 \caption{Self-supervised signal definitions (per channel)}

 \label{tab:signals}

 \centering

 \small

 \setlength{\tabcolsep}{4pt}

 \begin{tabular}{lp{5.5cm}}

  \toprule

  \textbf{Signal} & \textbf{Definition} \\

  \midrule

  EOT & Speech offset; other speaker takes floor within 4s \\

  HOLD & All other speech offsets (not EOT) \\

  BOT & Speech onset ($\geq$1s) following the other speaker \\

  BC  & Isolated utterance $\leq$1s, $\geq$1s silence before/after \\

  VAD & Binary voice activity per frame \\

  FVAD & Mean voice activity over [0--240ms], [240--480ms], [480--960ms], [960ms--2s] ahead \\

  \bottomrule

 \end{tabular}

\end{table}

We use a 4s lookahead (vs.\ VAP's 1s) to avoid discarding the 12\% of Switchboard turn transfers with pauses exceeding 1s. EOT and HOLD are complementary by construction: at every speech offset exactly one fires, giving the model the ability to discriminate turn ends from mid-turn pauses without any labeling. We apply focal loss~\cite{lin2017focal} to sparse signals and binary cross-entropy to VAD and FVAD. Labels are temporally smoothed with an asymmetric Gaussian ($\sigma{=}3$ frames before each event, $\sigma{=}1$ frame after), training the model to anticipate each signal up to 240ms early.

\subsection{Agent Action Inference}

The predicted signals (six per channel) are mapped to five agent actions (Table~\ref{tab:actions}). ST and CL correspond to VAP's shift and hold decisions, while SL, CT, and BC capture overlap and backchannel events that existing models leave unaddressed. As a zero-parameter baseline, domain-knowledge heuristics threshold individual signals directly (e.g.\ $\text{EOT}_\text{u} {>} 0.5 \wedge \text{BOT}_\text{a} {>} 0.5 \Rightarrow \text{ST}$). For richer class decisions from the predicted signals, we also use a multinomial logistic regression (LR) fitted on held-out validation data, that maps signal probabilities to agent actions. The coefficients of the learned LR are interpretable. For example, the start-talking action focuses more on $\text{EOT}_\text{u}$ and $\text{VAD}_\text{a}$ (user turn ended, agent is speaking), while negating $\text{VAD}_\text{u}$ and $\text{BOT}_\text{u}$ (user silent, not restarting). LR combines the voice activity, future activity, and event signals (EOT, BOT, HOLD, BC) into a single agent action.

\begin{table}[t]

 \caption{Agent actions and ground-truth definitions.}

 \label{tab:actions}

 \centering

 \footnotesize

 \setlength{\tabcolsep}{3pt}

 \begin{tabular}{llp{4.2cm}}

  \toprule

  \textbf{} & \textbf{Action} & \textbf{Definition} \\

  \midrule

  ST & Start-talking   & User offset; agent speaks within 4s \\

  CL & Continue-listening & User offset; user resumes within 2s \\

  SL & Start-listening  & Overlap onset; incoming speech $>$1s \\

  CT & Continue-talking  & Overlap onset; incoming speech $<$1s \\

  BC & Backchannel    & Agent vocalization $<$1s during user speech \\

  \bottomrule

 \end{tabular}

\end{table}

\section{Experiments}

\subsection{Setup}

We used approximately 453\,h of dual-channel conversation audio for pretraining the model. The dataset consists of otoSpeech~\cite{otospeech2025} (289\,h of 1,125 English full-duplex online conversations at 24kHz) and Switchboard~\cite{godfrey1992switchboard} (220\,h of telephone speech at 8kHz, with the 138-session test split held out for evaluation). In Stage-1, we trained the LoRA adapters~\cite{hu2022lora} with $\mathrm{lr}{=}3{\times}10^{-4}$ followed by full finetuning of LLM parameters with $\mathrm{lr}{=}10^{-5}$ and early stopping on a single A100 40GB GPU with a batch size of 32. In Stage-2, we only finetune the prediction heads and LoRA adapters with $\mathrm{lr}{=}1{\times}10^{-4}$ for up to 10 epochs, using early stopping and dropping the generative loss. The Mimi codec was frozen in both stages.

The Switchboard dataset is primarily used for evaluation, and the same 138-session test split used in prior works~\cite{ekstedt2022vap,wang2024turntaking} is evaluated. otoSpeech provides a secondary evaluation on a disjoint 113-session test split. External baselines are VAP~\cite{ekstedt2022vap} (5.8M parameters, CPC encoder, dual-channel audio) and the best published audio+text system~\cite{wang2024turntaking} (up to 3.1B parameters, RedPajama\,+\,HuBERT). Ablation variants A--G defined in Section~\ref{sec:ablation} are included in Table~\ref{tab:agent_action}.

Heuristics or the LR probe are used to infer agent actions from the signals predicted by the 12 signal heads (Section~2.4). The probe can only linearly aggregate signals or probabilities that the neural model already captures. The original paper definitions are used for evaluation on the VAP frame-level protocol~\cite{ekstedt2022vap} and Wang et al.~\cite{wang2024turntaking}, and the same Switchboard test set is used for evaluation. All models in Table~\ref{tab:agent_action} are evaluated under identical 4\,s action definitions. Table~\ref{tab:vap_protocol} further evaluates DualTurn under VAP's original 1\,s conditions to confirm that gains are not from the label definition but from learned representations.

\subsection{Main Results}

Wang et al.~\cite{wang2024turntaking} provide the most direct external comparison with word-level benchmarking. The learned representations are strong enough that even the single EOT signal with no aggregator (AUC avg 0.914) exceeds the 3.1B audio+text model of Wang et al.\ (AUC avg 0.880) without combining with other signals. The performance further improves to AUC avg 0.930 by combining multiple learned signals, and the LR probe reaches the best performance of AUC avg 0.963 (Table~\ref{tab:word_level}).

VAP (native) applies VAP's own $p_{\text{now}}$ and $p_{\text{future}}$ signals directly to agent action decisions and VAP (LR-6) fits a logistic regression over VAP's six projection outputs on the validation set. DualTurn outperformed VAP on all five agent action classes across both evaluation datasets (wF1 0.633 vs.\ 0.389 on Switchboard, wF1 0.707 vs.\ 0.461 on otoSpeech), Table~\ref{tab:agent_action}. Backchannel detection shows the largest gap, where VAP achieves BC\,F1\,=\,0.000 because it has no dedicated BC signal~\cite{inoue2025backchannel}. VAP cannot distinguish BC from CT even with an LR probe on its predicted projection probabilities. Trained entirely on self-supervised labels like VAP, DualTurn achieves BC\,F1\,=\,0.349 (chance F1\,$\approx$\,0.080) in predicting "the agent should backchannel now". This confirms the benefits of generative pretraining in Stage-1. The Stage-1 pretrained and unpretrained models show a clear binary split in BC\,F1 seen in Table~\ref{tab:agent_action}, and this is analyzed in Section~\ref{sec:analysis}.

DualTurn outperformed VAP on all four tasks (Table~\ref{tab:vap_protocol}), with Shift/Hold gaining the most ($+$0.142 wF1) and Short/Long, Shift-Prediction, and BC-Prediction showing consistent improvements on VAP's own evaluation protocol, despite being trained on our broader 4s definition.

\begin{figure}[t]

 \centering

 \includegraphics[width=\columnwidth]{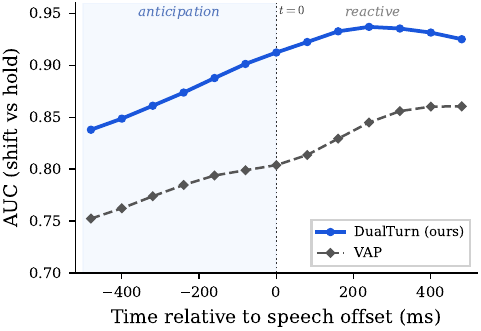}

 \caption{Shift-vs-Hold AUC at each time offset relative to speech offset (Switchboard test set).}

 \label{fig:anticipation}

\end{figure}

DualTurn picks up on turn boundaries about 220\,ms before VAP does, reacting at a median of $-360$\,ms vs.\ $-140$\,ms relative to turn end (Figure~\ref{fig:anticipation}). It is also less likely to misfire: ST-for-CL confusions drop from 27.4\% to 22.4\%, ST\,F1 improves to 0.829 over VAP's 0.808, and interruptions fall by 5 percentage points.

\begin{table}[t]

 \caption{Agent action prediction on Switchboard (138-session) and otoSpeech (113-session). wF1\,=\,weighted F1 (5 classes). Ant@$-$240\,=\,shift/hold AUC 240\,ms before offset (chance\,=\,0.500). Variants A--G defined in Section~\ref{sec:ablation}.}

 \label{tab:agent_action}

 \centering

 \small

 \begin{tabular}{l c c c}

  \toprule

  \textbf{Model} & \textbf{wF1} & \textbf{BC F1} & \textbf{Ant@$-$240} \\

  \midrule

  \multicolumn{4}{l}{\textit{--- Switchboard ---}} \\

  \midrule

  VAP (native) & .276 & --- & .785 \\

  VAP (LR-6) & .389 & .000 & .780 \\

  \midrule

  \textbf{A -- LoRA (main)} & \textbf{.633} & \textbf{.349} & .874 \\

  B -- LoRA + CB & .613 & .077 & .876 \\

  C -- No Pretrain & .604 & .079 & .863 \\

  D -- LSTM & .602 & .077 & .868 \\

  E -- Full FT & .626 & .337 & \textbf{.879} \\

  F -- Discrete & .602 & .072 & .867 \\

  G -- Text-Aware PT & .605 & .085 & .873 \\

  \midrule

  \multicolumn{4}{l}{\textit{--- otoSpeech ---}} \\

  \midrule

  VAP (LR-6) & .461 & .000 & .719 \\

  E -- Full FT & \textbf{.709} & .498 & \textbf{.860} \\

  A -- LoRA (main) & .707 & \textbf{.512} & .848 \\

  \bottomrule

 \end{tabular}

\end{table}

\begin{table}[t]

 \caption{Frame-level VAP protocol on Switchboard, evaluated under VAP's original 1s bidirectional event conditions~\cite{ekstedt2022vap}. All metrics: wF1. S/H\,=\,Shift/Hold, S/L\,=\,Short/Long, S-P\,=\,Shift-Prediction, BC-P\,=\,BC-Prediction. VAP paper used 11-fold CV (S/H\,=\,0.899); our fixed-split re-evaluation gives 0.843.}

 \label{tab:vap_protocol}

 \centering

 \small

 \begin{tabular}{l c c c c}

  \toprule

  \textbf{Model} & \textbf{S/H} & \textbf{S/L} & \textbf{S-P} & \textbf{BC-P} \\

  \midrule

  VAP (our eval) & .843 & .916 & .720 & .838 \\

  \midrule

  A -- LoRA & \textbf{.985} & .979 & .764 & .864 \\

  E -- Full FT & .982 & \textbf{.985} & \textbf{.771} & \textbf{.869} \\

  \bottomrule

 \end{tabular}

\end{table}

\begin{table}[t]

 \caption{Word-level turn prediction on Switchboard. C\,=\,continue, B\,=\,backchannel, T\,=\,turn-shift. Baselines from~\cite{wang2024turntaking}. A--EOT: single EOT signal, 0 params. A--Heur: multi-signal heuristic, 0 params. A--LR: LR probe over all signal heads.}

 \label{tab:word_level}

 \centering

 \footnotesize

 \setlength{\tabcolsep}{3.5pt}

 \begin{tabular}{l c c c c c}

  \toprule

  \textbf{Model} & \textbf{AUC(C)} & \textbf{AUC(B)} & \textbf{AUC(T)} & \textbf{Avg} & \textbf{EER} \\

  \midrule

  GPT-2 & .851 & .774 & .862 & .829 & 24.5 \\

  RP+HuBERT+hist & .903 & .818 & .920 & .880 & 19.3 \\

  \midrule

  A -- EOT only & .918 & .904 & .919 & .914 & 15.2 \\

  A -- Heuristic & .940 & .925 & .924 & .930 & 13.2 \\

  \textbf{A -- LR-probe} & \textbf{.961} & \textbf{.979} & \textbf{.950} & \textbf{.963} & \textbf{9.7} \\

  \bottomrule

 \end{tabular}

\end{table}

\subsection{Analysis}\label{sec:analysis}

\textbf{Speech Pretraining Switches BC Action Prediction On.} Table~\ref{tab:agent_action} shows a split in backchannel performance, with every model falling into one of two categories based on whether it was Stage-1 pretrained or not. While every other model is capped at BC\,F1 0.085, Models A and E, the only two with Stage-1 pretraining and a clean Stage-2 classification objective, achieve BC\,F1 of 0.337--0.349. The improvement in BC is substantial rather than incremental, with recall increasing from 0.045 to 0.458, corresponding to a relative gain of +340\% in F1. But the precision is only 0.282 (vs.\ chance 0.080) given the backchannels are less than 8\% of all events in the evaluation data, which shows the signal is real but modest. The pretraining makes it feasible to predict the agent backchannel action even with this sparsity, but does not solve it yet. When deployed, the BC signal shouldn’t act as the sole trigger. Instead, it should either operate with a higher threshold or be used as a soft input to guide the decision policy.

\textbf{Architecture Alone Cannot Reproduce This.} The performance is almost identical for the 8M LSTM model (D) and the 0.5B LLM (C), with wF1 0.602 vs.\ 0.604 and BC\,F1 0.077 vs.\ 0.079. Over 99\% of gains in total BC\,F1 come from pretraining (A vs.\ C), not from architecture (D vs.\ A). The contribution of the LLM is only about $+$0.002 BC\,F1. Without pretraining, the LLM's capacity advantage over the LSTM is negligible, but pretraining in turn requires sufficient capacity to be useful. The backbone is not the source of turn-taking knowledge but a necessary vessel that pretraining fills.

\textbf{Semantics and Prosody Drive Turn-Taking.} The Stage-2 turn-taking prediction signals fall into two difficulty classes. VAD and FVAD are dense per-frame signals learned easily by all variants, including the LSTM. BOT, BC, and to a lesser degree EOT are sparse interactional events, requiring recognition of two-speaker dynamics from limited training signal. This gap is closed by pretraining, with $+$188\% improvement for BOT and $+$41\% for BC, while VAD and FVAD show minimal improvements. We performed a codebook ablation on the trained discrete Model-F, by zeroing out or isolating individual codebooks. The results show a consistent ordering of contributions of codebooks on shift/hold AUC. 56\% of the turn-end discrimination signal comes from CB0 (semantics), CB1 contributes 26\%, and the remaining 18\% from CB2--7 together. The contribution of semantic and prosodic information is more than 80\% and the fine acoustics contribution is minimal. This is also consistent with prior findings~\cite{gravano2011turntaking}.

\textbf{Pretraining Creates Multi-Scale Attention Hierarchies.} We performed an attention analysis on the trained DualTurn model by probing the mean attended temporal distance of each head/layer of the LLM. It shows 6 layers of the LLM attend to short-range ($<$1s), probably handling frame-level acoustic detail, and 3 layers attend to long-range (L7\,=\,14.9s, L9\,=\,12.6s, L11\,=\,14.8s), maintaining conversational context at a temporal scale that is not achievable with the LSTM. This explains why C$\approx$D in performance: without Stage-1 pretraining, only 3 layers of Model~C attend to short-range, with scattered long-range attention. The pretrained transformer also develops action-specific behavior without explicit supervision. The user channel is attended 3.77$\times$ more for the CL class and 2.21$\times$ for the BC class. Each action class's temporal reach also differs: 3.41s for BC and 4.91s for CT.

\subsection{Ablation Study}\label{sec:ablation}

Table~\ref{tab:agent_action} includes seven variants: \textbf{A} (main; LoRA, pretrained, continuous Mimi, Transformer), \textbf{B} (retains the auxiliary generative loss in Stage-2), \textbf{C} (no Stage-1 pretraining), \textbf{D} (8M-parameter LSTM backbone), \textbf{E} (full fine-tuning of all 500M parameters), \textbf{F} (discrete codebook indices), and \textbf{G} (text generation ASR objective added to Stage-1).

The pretraining and backbone effects are discussed in Section~\ref{sec:analysis} (A vs.\ C, A vs.\ D vs.\ C). Even with 55$\times$ fewer parameters, LoRA (A) slightly beats full finetuning (E) on both BC\,F1 (0.349 vs.\ 0.337) and wF1 (0.633 vs.\ 0.626). We compared using discrete codebook indices (F) with continuous Mimi representations (A) as inputs with Stage-1 pretraining and found that the continuous input representations outperformed the discrete codebook inputs on all metrics (wF1 0.633 vs.\ 0.602, BC\,F1 0.349 vs.\ 0.072), as quantizing the continuous representation to discrete indices destroys the prosodic nuances, and the discrete model has to learn the codebook embeddings from scratch.

We also have two negative ablation design experiments. Adding Stage-1 generative loss as an auxiliary loss drops the BC\,F1 from 0.349 to 0.077 (A vs.\ B), and wF1 from 0.633 to 0.613, as competing gradients suppress the sparse task(EOT, BOT, HOLD, BC) learning. To make use of the LLM's pretrained text generation capability and force the parameters to learn speech-text modality adaptation, we added a text output objective (ASR task) in Stage-1 and observed that performance dropped in Stage-2 (A vs.\ G: wF1 0.633 vs.\ 0.605, BC\,F1 0.349 vs.\ 0.085), showing that better representations are learned for turn-taking tasks in audio-only modality training than with text-aligned pretraining.

\section{Conclusion}

Pretraining a dual-channel generative speech model on conversational audio and finetuning for explicit turn-taking signals narrows the gap between silence-based endpointing methods and S2S-level turn-taking dynamics. Without human annotations, DualTurn outperformed other methods like VAP and the 3.1B audio-text fusion model with better anticipatory behavior, predicting turn ends 220\,ms earlier than VAP. The multi-scale attention hierarchies enhanced by pretraining enabled the model to predict agent backchannels. Scaling to larger, multilingual/multi-party corpora and optimizing the generative pretraining are natural next steps, as the training and evaluation were limited to 453\,h of English two-party speech in our experiments. The main takeaway is that for turn-taking, generative pretraining is the teacher and the LLM backbone is the vessel, not the source of knowledge.

\bibliographystyle{IEEEtran}

\bibliography{mybib}

\end{document}